\documentstyle[preprint,aps]{revtex}
\begin{document}
\draft
\title{Laughlin liquid - Wigner solid transition \\
at high density in wide quantum wells}
\author{Rodney Price$^1$, Xuejun Zhu$^2$, S. Das Sarma$^1$, and
P.\ M.\ Platzman$^2$}
\address{$^1$Physics Dept., University of Maryland, College
Park, MD  20742 \\
$^2$AT\&T Bell Laboratories, Murray Hill, NJ 07974}
\maketitle

\begin{abstract}
Assuming that the phase transition between the Wigner solid and
the Laughlin liquid is first-order, we compare ground-state
energies to find features of the phase diagram at fixed $\nu$.
Rather than use the Coulomb interaction, we calculate the
effective interaction in a square quantum well, and fit the
results to a model interaction with length parameter $\lambda$
roughly proportional to the width of the well.  We find a
transition to the Wigner solid phase at high density in very
wide wells, driven by the softening of the interaction at short
distances, as well as the more well-known transition to the
Wigner solid at low density, driven by Landau-level mixing.
\end{abstract}

\narrowtext
\newpage

In the past few years, experiments with 2-dimensional electron
systems in a perpendicular magnetic field have shown a
long-expected behavior as the strength of the field increases.
At comparatively low fields, the now-familiar fractional quantum
Hall effect appears, as the longitudinal resistivity $\rho_{xx}$
falls exponentially to zero as the temperature $T \rightarrow 0$
at fractional filling factors $\nu$.  As the field increases,
however, an insulating state appears, with $\rho_{xx}
\rightarrow \infty$ when $T \rightarrow
0$.\cite{willett88,jiang90,goldman90}  Optical experiments
\cite{buhmann90,buhmann91} show a new spectral line developing
at the same fields.  This behavior is commonly thought to signal
the presence of a Wigner solid, predicted many years ago, and
pieces of an experimentally-derived phase diagram between the
Wigner solid and the fractional quantum Hall (FQHE) or Laughlin
liquid can be sketched out.

There are several experimental parameters that affect the
boundary between Wigner solid and FQHE
liquid.\cite{lam84,zhu93,price93,platzman93}  The most important
is the magnetic filling factor $\nu = 2 \pi n \ell^2$, where $n$
is the electron density and the magnetic length $\ell = (\hbar
c/eB)^{1/2}$.  The FQHE appears only at certain fractional
filling factors $\nu = p/q$, where $p,q$ are integers and $q$ is
odd.  The Wigner solid exhibits no such detailed dependence on
$\nu$, but becomes gradually more favorable as the particles are
localized with decreasing $\nu$.  Also important is the electron
density $n$, parameterized by the ion-disk radius $r_s = (\pi
n)^{-1/2}$.  (Here we use atomic units, where length is measured
in units of the Bohr radius $a_B = \hbar^2/me^2$ and energy in
units of $e^2/a_B$.)  As the Landau-level separation is $\hbar
\omega_c = 2/\nu r_s^2$, the energy cost of localizing the
particles by Landau-level mixing falls rapidly with increasing
$r_s$ at fixed $\nu$, until at some critical $r_s$ the Wigner
solid becomes more favorable than the FQHE liquid and the system
freezes.

Finally, the experiments are all done in real systems, which
must be considered quasi-two-dimensional, with some finite
thickness $L$ characterizing the width of the electron layer.
On average a pair of electrons is separated in the $z$-direction
by $\lambda < L$, so their effective interaction at distances $r
\ll \lambda$ in the $xy$ plane becomes much softer than the
Coulomb interaction, while at large distances $r \gg \lambda$,
the interaction is essentially Coulombic.  This preserves the
long-range character of the interaction while weakening the
short-range part.  Because the formation of the Wigner solid is
driven by the long-range part of the interaction, while the FQHE
liquid derives its energy advantage from the short-range part,
the quasi-two-dimensional character of a real experimental
system might be expected to favor the solid somewhat.  There
have been several recent experiments and theoretical studies
\cite{suen92,he90} of finite thickness effects on the
incompressible FQHE state and, in general, it is now
well-established that finite layer thickness tends to destroy
incompressibility by reducing the short-range part of the
Coulomb interaction.  To the best of our knowledge, however, the
enhancement of the Wigner solid phase by the finite thickness
effect has not been examined theoretically.

In this paper, we study the quantitative effects of the three
length scales $\ell$, $r_s$, and particularly $\lambda$ on the
Laughlin liquid - Wigner solid phase diagram. We compare
ground-state energies and excited states of variational
wavefunctions for the Wigner solid and the FQHE liquid as both
$r_s$ and $\lambda$ are allowed to vary.  We find that at small
$\lambda$, such as that found in GaAs heterojunctions, the
zero-temperature phase transition from liquid to solid at large
$r_s$ does not vary much from the Coulomb case.\cite{price93}
At large $\lambda$ we find that the FQHE liquid gives way to the
Wigner solid at low density, as expected, but most unexpectedly,
we find that the Wigner solid phase dominates the FQHE liquid at
high density as well.  Only in an intermediate range of $r_s$
does the liquid have lower energy than the solid when $\lambda$
is large.  The interplay among the length scales $\ell$, $r_s$,
and $\lambda$ can, in principle, therefore lead to a reentrant
Wigner solid transition at large $\lambda$ as $r_s$ is varied.

The electrons in a quantum well are confined to a small number
of subbands, usually just one, and then can be thought of as
extended rod-like charges in the $z$-direction which are allowed
to move only in the $xy$ plane.\cite{ando82}  This approximation
has been used in the past to study the weakening and eventual
collapse of the FQHE state in a wide quantum well.\cite{he90}
The interaction between these model charges is given by
\begin{equation}
V_{\rm eff} = \int dz_1 \int dz_2 \frac{|\zeta(z_1)|^2
|\zeta(z_2)|^2}{\left[ r^2 + (z_1-z_2)^2 \right]^{1/2}}.
\label{effective_interaction}
\end{equation}
Here $\zeta(z)$ is the envelope wavefunction describing
quantization in the $z$-direction and $r$ is the separation
between electrons in the $xy$ plane.

The confining potential in the $z$-direction enters into this
equation only through the envelope wavefunction $\zeta(z)$.
This wavefunction should be obtained in a self-consistent
procedure which takes into account the interaction of the
electrons in the $xy$ plane.  At low electron density, in
quantum wells where the subband splitting is large, $\zeta(z)$
is simply the $z$-component of the single-particle wavefunction.
 When electron density becomes higher, or the well is made wider
bringing the subbands closer together, $\zeta(z)$ will be
modified.

In a square well, however, $\lambda$ and $r_s$ are (roughly
speaking) independent parameters, as the width of the well $L$
is to lowest order independent of the density of the electrons
in the $xy$ plane.  It is only when subband mixing begins to
become important that the electron density begins to affect the
envelope wavefunctions $\zeta(z)$.  We expect that when only the
lowest subband is occupied the single-particle wavefunctions
will be adequate for $\zeta(z)$. The usual cosine solution for
an infinite well is shown in the inset to Figure \ref{hswell},
as well as a gaussian wavefunction
\begin{equation}
\zeta_{\rm sq}(z) = \frac{1}{(\pi \gamma^2)^{1/4}} e^{-z^2/2
\gamma^2},
\end{equation}
fitted to the cosine solution.  Here $\gamma = 0.277 L$ gives
the gaussian wavefunction shown in Figure \ref{hswell}.  The
effective potential for the two wavefunctions is almost
identical.  The gaussian wavefunction gives
\begin{equation}
V_{\rm eff}^{\rm sq}(r) = \frac{1}{\sqrt{2 \pi} \gamma}
e^{r^2/4\gamma^2} K_0\left(\frac{r^2}{4 \gamma^2}\right),
\end{equation}
where $K_0$ is a modified Bessel function.  We would like to use
the simpler model interaction of Zhang and Das
Sarma,\cite{zhang86}
\begin{equation}
V_0(r) = \frac{1}{\sqrt{r^2+\lambda^2}},
\label{modelint}
\end{equation}
so we choose the parameter $\lambda$ to fit $V_{\rm eff}^{\rm
sq}(r)$ best when $r$ is large.  A least-squares fit, shown in
Figure \ref{hswell}, yields $\lambda/L = 0.2$.

Because $\lambda/L$ is small for the square well, we need a wide
well if we are to investigate a system with reasonably large
$\lambda$.  For example, a well with $\lambda = 1$ in an
electron system in GaAs is approximately 500\AA wide.  In these
wide wells, subband mixing can become important, and the
envelope wavefunctions will tend to spread out toward the edges
of the well as the electrons reduce their potential energy.  We
have used the self-consistent approach taken in \cite{suen92} to
estimate the density distribution $n(z)$, and we find that
$\lambda/L$ will vary anywhere from 0.15 to 0.30 in the presence
of subband mixing.\cite{pablo}  If we choose the lowest subband
value $\lambda = 0.2L$ for the square well, $V_0(r)$ is nearly
identical to $V_{\rm eff}^{\rm sq}(r)$ for $r \ge 0.2L$, and
differs significantly from $V_{\rm eff}^{\rm sq}(r)$ only for $r
\le 0.1L$.  The pair correlation function for both liquid and
solid is small below about $r = r_s$, so we believe $V_0(r)$ to
be a good description of $V_{\rm eff}^{\rm sq}(r)$ for $\lambda
\le 2$.

In extremely wide quantum wells, $n(z)$ will be modified further
as it peaks near the well edges.  The well then begins to
resemble a highly-coupled double-layer system.  Electrons in the
well may lower their potential energy by localizing near the
edges of the well, at the cost of some kinetic energy.  In this
more complicated situation, the approximation
(\ref{effective_interaction}) begins to break down.  At high
density, the interaction between electrons that are adjacent in
the $xy$-plane becomes small, as they become separated in the
$z$-direction.  The interaction between electrons localized on
the same side of the well becomes stronger, however, and the net
effect is a small potential energy savings.  We should note,
however, that the magnetic field will tend to suppress subband
mixing, as at $\nu = 1/3$ the filling in the lowest subband will
be at most 1/3.


In this paper, we determine the zero-temperature phase boundary
between Wigner solid and Laughlin liquid by comparing
ground-state energies of variational wavefunctions for the
liquid and the solid.  Because we are simply comparing energies,
we are assuming that the phase transition is first-order, and we
neglect the possible presence of any other states in the
vicinity.  At high electron density we regard the approximation
(\ref{effective_interaction}) as a qualitative guide only, and
do not attempt to predict a critical $r_s$ and $\lambda$
quantitatively for the low $r_s$ transition.

A variational wavefunction for the liquid which interpolates in
some sense between a wavefunction with the lowest possible
kinetic energy, the Laughlin wavefunction, and a wavefunction
with the lowest possible potential energy, in which the
electrons are completely localized, might be expected to be a
good variational choice.  To that end, we have chosen a
variational wavefunction that consists of the Laughlin
wavefunction $\psi_m$ ($m = 1/\nu$) multiplied by a Jastrow
factor $\prod_{i<j} e^{-\alpha/\sqrt{r_{ij}}}$, where $r_{ij}$
is the distance between the $i^{\rm th}$ and $j^{\rm th}$
particles and $\alpha$ is the variational parameter.  When
$\alpha = 0$, we recover the Laughlin wavefunction, and when
$\alpha \neq 0$, the wavefunction is no longer analytic and
higher Landau levels are mixed in.  The Jastrow factor
introduces more correlations into the wavefunction, lowering the
potential energy, while introducing a kinetic energy cost.

Details of the calculation are given in Ref.\ref{price93}, so we
will only review it briefly here.  (Note that \cite{price93}
uses units of energy $e^2/2a_B$, while this paper uses atomic
units $e^2/a_B$.)  We use the spherical geometry, in which our
wavefunction becomes
\begin{equation}
\psi^\alpha_m = \prod_{i<j} (u_i v_j - u_j v_i)^m
	\exp\left(\frac{-\alpha}{|u_i v_j - u_j v_i|^{1/2}}\right),
\label{laughlin-jastrow}
\end{equation}
where $ u_i \equiv e^{-i \phi_i/2} \cos(\theta_i/2), v_i \equiv
e^{i \phi_i/2} \sin(\theta_i/2)$ are convenient spinor
coordinates, and the distance between particles $i$ and $j$ is
taken as the chord distance $r_{ij} = 2R|u_i v_j - u_j v_i|$.
Evaluating the energy of this wavefunction by Monte Carlo and
minimizing at a fixed $r_s$ and $\lambda$ gives the results
shown by the dashed lines in Figures \ref{energy_lambda} and
\ref{energy_rs}.

In order to find the liquid-solid phase boundary we need a
rather accurate evaluation of the solid.  Lam and Girvin
\cite{lam84} evaluated the energy of a correlated Wigner solid
wavefunction
\begin{equation}
\Psi = \exp \left( \frac{1}{4} {\sum_{i,j}}^\prime \xi_i B_{ij}
\xi_j \right) \prod_i \phi_{R_i}(z_i),
\end{equation}
where $\xi_i = z_i - R_i$, $B_{ij} \equiv B(R_i - R_j)$, and
\begin{equation}
\phi_{R_i}(z_i) = \exp \left( -\frac{1}{4} \left[ | z_i - R_i |
^2 - (z_i^* R_i - z_i R_i^*) \right] \right).
\label{single_particle_wf}
\end{equation}
Here $z_i = x_i + i y_i$ is the $i^{\rm th}$ particle position
and $R_i = X_i + i Y_i$ is the $i^{\rm th}$ lattice site.
$\Psi$ is the harmonic crystal wavefunction restricted to the
lowest Landau level, and the variational parameters $B_{ij}$ are
calculated by using the values derived from the harmonic
crystal.  However, in order to make a reasonable comparison of
solid and liquid wavefunctions, we need, as discussed above, a
wavefunction which includes Landau-level mixing.

A study \cite{zhu93} of the ground-state energy of the Wigner
crystal including Landau-level mixing, has recently been
completed, using the Coulomb interaction.  We have extended
their work to use the modified potential (\ref{modelint}).  This
calculation is similar to Lam and Girvin's, except that two more
variational parameters $\alpha$ and $\beta$ are added to the
wavefunction to put in more correlations at the expense of some
Landau-level mixing.  First, the gaussians in the
single-particle wavefunctions (\ref{single_particle_wf}) were
``squeezed'' to move the electrons farther away from each other,
by making the replacement
\begin{equation}
\exp \left( -\frac{1}{4} |z_i - R_i|^2 \right) \rightarrow \exp
\left(-\beta|z_i - R_i|^2 \right).
\end{equation}
Varying the parameter $\beta$ away from 1/4 introduces
Landau-level mixing into the wavefunction because the
single-particle wavefunctions $\phi^\beta_{R_i}(z_i)$ are no
longer eigenstates of the single-particle Hamiltonian.  An
additional Jastrow factor is then introduced, and the final
wavefunction, with two variational parameters $\alpha$ and
$\beta$, is then
\begin{equation}
\Psi = \exp \left( {\sum_{i,j}}^\prime \left[ \frac{1}{4} \xi_i
B_{ij} \xi_j - \frac{\alpha}{2} u(|z_i - z_j|) \right] \right)
\prod_i \phi^\beta_{R_i}(z_i),
\end{equation}
where
\begin{equation}
u(r) = \frac{1}{\sqrt{r}} \left( 1 - e^{-\sqrt{r/F} - r/2F}
\right),
\end{equation}
and $F$ is a constant chosen to optimize the pseudo-potential at
small $r$.  Zhu and Louie varied $B_{ij}$ as well as $\beta$,
but found that varying $B_{ij}$ had very little effect on the
energy, as Lam and Girvin had suggested.  The quantum Monte
Carlo calculations were done with modified periodic boundary
conditions,\cite{zhu93} which require the addition of a phase
factor to the single-particle wavefunctions
(\ref{single_particle_wf}).  The results are shown as the solid
lines in Figure \ref{energy_lambda} and the triangular data
points in Figure \ref{energy_rs}.

The $r_s = 2$ curves in Figure \ref{energy_lambda} are very
nearly the lowest Landau-level energies, since $\hbar \omega_c$
is large at $r_s = 2$.  In fact, if each of the curves in Figure
\ref{energy_lambda} were plotted as a function of $\lambda/r_s$
and there were no Landau-level mixing, they would lie on top of
each other.  Only the increased Landau-level mixing at $r_s =
10$ and 20 reduces the energy there somewhat.  The parameter
$\lambda/r_s$ can be thought of as the ratio of the average
separation in the $z$-direction $\lambda$ and the average
separation in the $xy$ plane $r_s$ (actually $\sim 2 r_s$) of
two nearby electrons.  The lowest Landau level energies cross at
$\lambda = 6$ at $r_s = 10$ and $\lambda = 12$ at $r_s = 20$,
but the variational energies here predict a freezing transition
at $\lambda \approx 4$ at $r_s = 10$ and possibly $\lambda
\approx 6$ -- 7 at $r_s = 20$.  These results illustrate the
fact that Landau-level mixing does in general tend to favor the
solid somewhat.

The variational wavefunctions use the fact that localizing the
electrons by mixing in higher Landau levels will keep
nearest-neighbor electrons farther apart, and since the Coulomb
potential rises rapidly at small $r$ the energy savings can be
significant.  As the interaction softens, however, the energy
savings becomes small, and as a result the liquid and solid
energies at $r_s = 2$ are nearly identical to the lowest Landau
level energies.  Ortiz, Ceperley, and Martin \cite{ortiz93} have
recently used their fixed-phase quantum Monte Carlo method to
improve the liquid energy shown here somewhat.  The difference
in energy is significant at $r_s = 10$ and 20, but at $r_s = 2$
the change in energy for the Coulomb interaction is only $\sim
0.006$, a shift on the order of 10\%.

The rapid descent of the solid energy with $\lambda$ leads to a
freezing transition at about $\lambda = 1.2$ at $r_s = 2$,
brought on by the softening of the short-range part of the
interaction.  Keeping the well width constant and lowering the
density gives a smaller $\lambda/r_s$, and the system melts once
again until the increased Landau-level mixing at high $r_s$
causes the system to freeze again.  Figure \ref{energy_rs} shows
the energies of solid and liquid when $\lambda$ is fixed and
$r_s$ is allowed to vary.  At $\lambda = 0.5$ we find no
significant lowering of the solid energy at low $r_s$, but at
$\lambda = 1$ the solid energy at $r_s = 2$ is nearly as low as
the liquid energy, and at $\lambda = 2$ and 3 the solid energy
at $r_s = 2$ is much lower than the liquid energy.  In all cases
the liquid becomes favorable at lower density until Landau-level
mixing again causes the system to freeze.  The transition at
high density has been seen in the experiment of Suen {\it et
al},\cite{suen92}  where an insulating phase was observed in an
800\AA well at $\nu = 1/3$, $r_s = 1.7$, which became a
well-defined FQHE state when the density was lowered to $r_s =
2.2$.

We can contrast the wide well (i.e. large $\lambda$) situation,
which we have argued in this paper may be favorable to the solid
phase both at large and smaller $r_s$, with the situation of a
heterojunction with a metal gate which screens the
inter-electron Coulomb interaction, changing it to $V_{\rm eff}
(r \rightarrow \infty) \sim 1/r^3$ and $V_{\rm eff} (r
\rightarrow 0) \sim 1/r$.  For the gated heterojunction case
clearly the Laughlin liquid will be preferred because the
effective interaction remains Coulombic for small $r$.  Thus,
careful experiments in wide wells contrasted with those in gated
heterojunctions would go a long way in establishing the
liquid-solid phase boundary in strong-field 2D systems.

The authors wish to thank P.\ I.\ Tamborenea and Song He for
helpful discussions, and P.\ I.\ Tamborenea for providing us
with his data.  This work was supported by the US-ARO, the
US-ONR, and NSF.

\begin{figure}
\caption{The effective interaction (solid line), Coulomb
interaction (dash-dot line), and model interaction (dashed line)
for a square well of width $L$.  The model interaction parameter
$\lambda = 0.2L$.  The lowest subband cosine wavefunction and
gaussian fit are shown in the inset.}
\label{hswell}
\end{figure}

\begin{figure}
\caption{The variational liquid energy (dashed line) and solid
energy (solid line) shown after subtracting $\hbar \omega_c/2$
and the Madelung energy for various fixed values of $r_s$.}
\label{energy_lambda}
\end{figure}

\begin{figure}
\caption{The variational liquid energy (dashed line) and solid
energy (triangles) shown after subtracting $\hbar \omega_c/2$
and the Madelung energy for various fixed values of $\lambda$.}
\label{energy_rs}
\end{figure}

\end{document}